\documentclass{IEEEmce}

\usepackage[colorlinks,urlcolor=blue,linkcolor=blue,citecolor=blue]{hyperref}

\usepackage{framed}
\usepackage{textpos}

\jvol{22}
\jnum{2}
\jmonth{Mar/Apr}
\publisheddate{02 04 2024}
\currentdate{03 04 2024}
\jname{Authors' version - official version published in IEEE Security \& Privacy Magazine}
\pubyear{2024}
\doiinfo{(10.1109/MSEC.2024.3363829 (official version)}

\begin{document}


\title{Privacy Engineering From Principles to Practice:\\A Roadmap}

\author{Frank Pallas}
\affil{Paris Lodron University of Salzburg / TU Berlin}

\author{Katharina Koerner}
\affil{Daiki}

\author{Isabel Barberá}
\affil{Rhite}

\author{Jaap-Henk Hoepman}
\affil{Radboud University}

\author{Meiko Jensen}
\affil{Karlstad University}

\author{Nandita Rao Narla}
\affil{DoorDash}

\author{Nikita Samarin}
\affil{University of California, Berkeley}

\author{Max-R. Ulbricht}
\affil{Berlin Commissioner for Data Protection and Freedom of Information}

\author{Isabel Wagner }
\affil{University of Basel}

\author{Kim Wuyts}
\affil{PriceWaterhouseCoopers Belgium}

\author{Christian Zimmermann}
\affil{Robert Bosch GmbH}

\begin{abstract}
Privacy engineering is gaining momentum in industry and academia alike. So far, manifold low-level primitives and higher-level methods and strategies have successfully been established. Still, fostering adoption in real-world information systems calls for additional aspects to be consciously considered in research and practice.
\end{abstract}

\maketitle

\begin{textblock*}{\textwidth}(0cm,-17cm) 
\begin{center}
\setlength{\FrameSep}{0.3cm}
\begin{framed}
    \textit{Authors' version before final copy-editing by IEEE Security \& Privacy. \copyright2024 IEEE. The official version of record can be found at \url{https://ieeexplore.ieee.org/document/10488836}}
\end{framed}
        
\end{center}
\end{textblock*}

\enlargethispage{10pt}

\chapterinitial{With organizations facing} increasingly stringent data protection regulations and digital trust being at the heart of growing user expectations, privacy engineering is gaining traction as a distinct discipline in business and academia alike. Large enterprises are establishing dedicated privacy engineering departments and more and more scientific venues adopt privacy engineering as one of their central themes, confirming Lea Kissner’s and Lorrie Cranor’s 2021 designation of privacy engineers as the “superheroes” of the privacy profession.\footnote{Kissner, L., and Cranor, L. (2021). Privacy engineering superheroes. Communications of the ACM, 64(11), 23-25. \url{https://doi.org/10.1145/3486631}} 

Privacy engineering leverages concepts from disciplines as diverse as information security, jurisprudence, economics, and psychology to facilitate the development of systems that are privacy-friendly by design. It explicitly takes a comprehensive view of such systems and services as well as their development and socio-technical surroundings. This helps bridge the gap between practical implementations and traditional privacy and security research. It allows companies to better and more reliably comply with increasing enforcement of regulations such as the European General Data Protection Regulation (GDPR) or the California Privacy Rights Act (CPRA). Privacy engineering also helps companies increase trust in their data handling practices on the side of their customers, employees and business partners, as well as to demonstrate accountability to data protection authorities.

While a vibrant community of academic researchers and corporate privacy engineers have been progressing the field significantly during the last years, uptake in the industry at large is still relatively low. Even though numerous design methods and frameworks have been established – from privacy threat modeling frameworks such as LINDDUN\footnote{Wuyts, K., Sion, L., and Joosen, W. (2020, September). Linddun go: A lightweight approach to privacy threat modeling. In 2020 IEEE European Symposium on Security and Privacy Workshops (EuroS\&PW) (pp. 302-309). \url{https://doi.org/10.1109/EuroSPW51379.2020.00047} See also \url{https://linddun.org/}} to generic privacy design strategies\footnote{Hoepman, J. H. (2018). Privacy design strategies (the little blue book). \url{https://www.cs.ru.nl/~jhh/publications/pds-booklet.pdf}} – privacy is, broadly speaking and with the exception proving the rule, still no first-class member of modern, real-world information systems engineering.

Privacy engineering can do better. Striving towards an enhanced uptake of privacy engineering in practice, this article highlights key aspects that need to be emphasized more prominently in the discourse. Drawing from lessons learned in various research projects and from extensive industry experience, we want to shed light on underrepresented, albeit crucial aspects of privacy engineering in the context of modern information systems engineering, thereby fostering its wide adoption in practice.

\section{What, then, is privacy engineering?}

In our quest to unravel the core of privacy engineering, it becomes apparent that even the underlying concept of privacy is  – like fairness, art, or democracy – an “essentially contested” one: we can basically agree on a term and its desirability, but its actual meaning is subject to a broad variety of different interpretations and inherently eludes reaching broad consensus on a single definition.\footnote{Mulligan, D. K., Koopman, C., and Doty, N. (2016). Privacy is an essentially contested concept: a multi-dimensional analytic for mapping privacy. Philosophical Transactions of the Royal Society A: Mathematical, Physical and Engineering Sciences, 374(2083). \url{https://doi.org/10.1098/rsta.2016.0118}} The same is true for privacy engineering: Conceptions range from the design and implementation of anonymity-preserving algorithms and protocols to higher-order ones taking up methods and practices from software engineering, physical architecture, human-computer-interaction, or socio-technical systems design. 

To this existing spectrum, we want to add another point of view that puts an \emph{explicit emphasis on practical applicability in real-world information systems.} In particular, we look at privacy engineering from the perspective of enterprise information systems and architectures, established paradigms and practices for their development and operation in practice, and the associated requirements and constraints. By bringing these aspects into focus, we can identify and highlight the gaps that exist between the current state of the privacy engineering discourse and the prevailing practices within the realm of enterprise information systems. 

This, in turn, allows us to identify aspects of crucial importance for privacy engineering to better align with real-world information systems engineering and, thus, to increase its practical relevance, applicability, and adoption. In this regard, we do in the following particularly highlight the needs to 1) broaden the view beyond anonymization, data minimization and security, to 2) consciously recognize what we call “second-order non-functional properties” of privacy mechanisms, and to 3) relax on so far predominant “all-or-nothing” aspirations. Lastly, we also highlight 4) how the provision of technical artifacts that are easily re-usable in real-world environments can induce “indirected implementation obligations” and thereby foster the broad application of novel privacy mechanisms in practice.

\section{Broadening the view beyond anonymization, data minimization, and security}

While privacy engineering is often considered as merely an approach to implement anonymization and pseudonymization techniques or to ensure confidentiality\footnote{See, e.g., Iwaya, L. H., Babar, M. A., and Rashid, A. (2023). Privacy Engineering in the Wild: Understanding the Practitioners' Mindset, Organisational Aspects, and Current Practices. IEEE Transactions on Software Engineering. \url{https://doi.org/10.1109/TSE.2023.3290237}}, privacy engineering entails a much broader range of goals and activities. Privacy-related regulations, such as the GDPR or the CPRA, and non-regulatory frameworks, such as the Fair Information Practice Principles (FIPPs) or the OECD Privacy Principles, clearly call for further principles to be properly reflected in the design and implementation of real-world information systems. These principles include:
\begin{itemize}
    \item \textbf{Lawfulness (incl. legal basis such as consent):} The collection and processing of personal information has to be done in a lawful and fair manner. Under the GDPR and other privacy legislations this can mean that any collection or processing of personal data is to be considered unlawful unless properly legitimized. Beyond individual consent, which is quite prominent in academic discussions, this legitimation can also rest on other legal bases, e.g.,the necessity for fulfilling a contract (think of address data being processed by an online shop) or legal obligations (e.g., an employer forwarding income data to tax authorities). Technical approaches for interlinking collection and processing of personal data with the respective underlying legitimation (allowing for subsequent reviews whether they are still valid, for instance) are, however, largely lacking.
    \item \textbf{Purpose Limitation:} Slightly simplified, the principle of purpose limitation says that personal data are only to be processed for the purpose(s) they were initially collected for. For technically materializing this principle, information systems and the underlying data management solutions must allow to control the flow and use of personal data based on respective purposes and, thus, be  “purpose-aware” by design. Approaches for, e.g., purpose-based access control will certainly prove valuable here.
    \item \textbf{Data Minimization (incl. necessity):} Minimizing the amount of personal data being processed to what is absolutely necessary is what widespread “privacy” technologies for anonymization, pseudonymization, etc. are typically aimed at. Noteworthily, data minimization does not necessarily require to minimize the amount of data in general but only the amount of personal data. This can – albeit with some pitfalls – also be achieved by means of sufficiently reducing/removing the linkability between data and its subject". Similarly, in many cases, even a simple process for recording and maintaining data retention periods would already significantly limit the amount of personal data kept by many services in common use today.
    \item \textbf{Transparency:} To allow data subjects (users) to act and decide in a well-informed, self-sovereign manner, they must be provided with sufficient information on how their data is processed, for which purposes, etc. All this information needs to be provided in a way that users can access and understand based on their individual abilities. Today, it is typically provided in textual privacy policies which are, however, barely legible by laypersons and more often than not conflict with today’s well-established agile principles and practices of systems engineering. This is calling for more appropriate, technically mediated approaches and expecting industry to pick up state of the art approaches such as code scanning for personal information processing, utilizing APIs for communicating privacy policies of microservices, or alternative novel but mature transparency by design measures.
    \item \textbf{Security:} The traditional C-I-A triad (confidentiality, integrity, availability) of information security is also highly relevant in the context of privacy. Personal data needs to be kept confidential and integrity and availability of personal data are of crucial importance for avoiding any mistreatment (imagine, for example, unauthorized changes to or deletions of personal health records) as long as the data is actually relevant (while in case of irrelevance, the principle of data minimization would apply and explicitly call for deletion).
    \item \textbf{Accountability:} Like any other rule, privacy-related obligations would be rather meaningless without appropriate means for monitoring (or demonstrating) their fulfillment and for holding responsible parties accountable. With regard to privacy, this is traditionally achieved through a mix of technical and non-technical approaches ranging from well-documented systems architectures over various technical mechanisms for trustworthy computing to in-depth on-site inspections by authorities and certification auditors. Under current givens of often cross-organizational processing of personal data in highly distributed and continuously changing information systems, however, these established means do hardly suffice to appropriately ensure accountability anymore. 
\end{itemize}

Beyond these, further principles such as data portability (allowing data subjects to transfer data from one service provider to another) or accuracy and fairness (ensuring that data is actually correct, not biased, and can be reviewed, corrected or amended) may also be added to the set of relevant privacy principles that need to be reflected technically. Last but not least, non-regulatory conceptions of privacy also refer to similar principles that cannot be properly addressed by means of anonymization and security alone.

Instead of largely concentrating on ever-new anonymization and security techniques, practice thus calls for a more encompassing set of functionalities covering all above-mentioned principles. The technology scope of privacy engineering should thus be consciously broadened. Mapping above-mentioned principles to privacy-focused protection goals also including unlinkability, intervenability, and transparency (as, for instance, done in the “Standard Data Protection Model” proposed by German data protection authorities\footnote{See \url{https://www.bfdi.bund.de/EN/Fachthemen/Inhalte/Technik/SDM.html}}) may also prove valuable here.

\section{Recognizing functional and non-functional properties of privacy mechanisms (and acknowledging the importance of the latter)}

In information systems engineering, it is typically distinguished between functional and non-functional properties that systems have and respective requirements they must fulfill. Functional properties here refer to the core functionalities a system is meant to provide: a database stores and allows to query data or a travel planning service is able to calculate appropriate routes and travel times for different means of transportation. Non-functional properties or “qualities”, in turn, refer to “constraint[s] on the manner in which [a] system implements and delivers its functionality”.\footnote{Taylor, R., Medvidovic, N., and Dashofy, E.: Software Architecture: Foundations, Theory, and Practice, Taylor \& Francis, 2009, p. 447.} Performance, scalability, or even security and privacy are typically mentioned here. Such non-functional properties are often crucial for the practical applicability or adoption of a technical system or component, irrespectively of its capacity to fulfill functional ones.

For privacy technologies, in turn, a similar differentiation must be made. From this perspective, functional properties refer to the privacy functionality a technical artifact provides: a certain property-preserving encryption scheme allows for a well-defined set of operations to be executed on encrypted data, a purpose-based access control scheme allows to technically enforce the privacy principle of purpose limitation, and so on. This is what we typically find in technical papers presenting novel privacy mechanisms, protocols, etc.

Non-functional properties of respective technical artifacts are, however, only rarely discussed. Nonetheless, these are of crucial importance for achieving applicability in practice. Based on existing research, we can identify at least the following non-functional properties of privacy mechanisms to be decisive for their practical application, albeit only rarely discussed in the privacy engineering literature:
\begin{itemize}
    \item \textbf{Usability in relevant real-world information systems contexts:} One of the core requirements for privacy mechanisms to be actually adopted in practice is that they are provided as an easily (re-) usable artifact (e.g. library, package, component) that can be applied in conjunction with different systems of a particular class (e.g., different SQL databases) actually employed in practice. 
    \item \textbf{Coherent integration into established software stacks, architectures and development practices:} To foster practical adoption, a technical privacy mechanism must pay appropriate regard to the context it shall be applied in. A database extension with a modified query language, for instance, will hardly be applicable in conjunction with abstraction layers such as ORMs widely used in practice. Development paradigms and practices such as agile DevOps might also call for explicit recognition in the design of certain privacy mechanisms.\footnote{Gürses, S., and van Hoboken, J. (2017). Privacy after the agile turn. \url{https://doi.org/10.31235/osf.io/9gy73}} Aligning privacy engineering approaches with security practices, which are already much more mature and adopted in practice, would be another useful angle to ensure integration.
    \item \textbf{Developer-friendliness and low implementation efforts:} If a new technical privacy mechanism places a significant burden on the developers who shall apply or integrate it into their systems, this will hinder its adoption in a multitude of ways. Conversely, if applying a privacy mechanism merely requires minimal code modifications, developers will be way less reluctant. Similarly, management support also strongly depends on the implementation overheads that are to be expected. 
    \item \textbf{Reasonable and experimentally determined performance overheads in realistic settings:} In many cases, the performance overhead raised by a novel privacy mechanism is rather unknown. In practice, however, the overhead to be expected is of crucial importance for deciding about a privacy mechanism’s application. Explicitly provided overheads empirically gathered in experiments resembling real-world systems, environments, and workloads as closely as possible are therefore indispensable for making conscious and empirically well-founded decisions. 
\end{itemize}

In the light of the above-mentioned conception of privacy itself being a non-functional property of information systems, we refer to these properties of privacy mechanisms as “second-order non-functional properties”. These (and presumably further ones) will foreseeably be decisive for a technical privacy artifact’s actual transfer from its scientific birthplace into real-world applications. Nonetheless, they are only marginally present in the privacy engineering discourse. 

\section{Let perfection not be the enemy of the good}

Another aspect quite prominent in the prevailing discourse regards the perceived need for solutions that provide some sort of formal guarantee that a given privacy property is 100\% ensured in the light of a certain attacker model. Of course, technical mechanisms able to achieve this would always be the first choice, but in many cases, these come at the cost of significant drawbacks in matters of practical applicability. Mechanisms for fully homomorphic encryption or secure multiparty computation are a prime example here: in theory, they allow to outsource critical calculations to external parties (such as cloud providers) while still providing confidentiality or integrity guarantees against these. However, such mechanisms usually come with tremendous performance overheads and lack easy integrability into real-world systems, hindering their application in practice. Similarly, adapted databases providing low-layer purpose-based access control have been proposed for materializing the principle of purpose limitation technically. However, these do not align with implementation stacks and data access models used in real-world information systems engineering, significantly limiting their practical applicability. Compared to these, alternative approaches for purpose-based access control explicitly aligned with such givens from practice while relaxing on aspects such as circumventability by adversarial in-house developers\footnote{See, e.g., Pallas, F., Ulbricht, M.R., Tai, S., Peikert, T., Reppenhagen, M., Wenzel, D., Wille, P. and Wolf, K., 2020. Towards application-layer purpose-based access control. In Proceedings of the 35th Annual ACM Symposium on Applied Computing (pp. 1288-1296). \url{https://doi.org/10.1145/3341105.3375764}} may turn out as the superior ones, given that they allow purpose-awareness to make it into real-world information systems at all. In matters of accountability, evidence doesn’t need to be “provably unforgeable” to provide an actual benefit, and so forth.

By and large, it needs to be better recognized that regulations do not require the implementation of  technical mechanisms that enforce privacy principles in a guarantee-like, 100\% fashion. Instead, they follow a non-binary, risk-based approach, calling for technical measures that properly reduce relevant risks (but not necessarily eliminate them completely and provably). The GDPR, for instance, obligates data controllers to apply technical and organizational measures “designed to implement [privacy] principles” and explicitly links respective obligations to factors such as the cost of implementation or the risks associated with the processing. From this perspective, an easy-to-implement, low-overhead mechanism that leaves a certain risk of circumvention by adversarial in-house developers can in many cases be preferable over one that provides formal guarantees, albeit at the cost of significant performance overheads. 

In consequence, privacy engineering should, more often than currently, take a “realistic stance” on developers and data controllers: It must be weighed whether it is more important and valuable to support \emph{them} in fulfilling their duties than trying to ensure absolute tamper- or concealment-proofness and end up without any mechanism being present at all. 

\section{Creating impact by shaping the state of the art}

One important question remains to be answered: how to foster the actual adoption of privacy engineering in the industry? Privacy engineering and privacy-friendly systems will almost always lead to increased development and operational costs. Hence, beyond their need to comply with legislation, data controllers often have only limited incentives to make their systems more privacy-friendly than absolutely necessary. 

Thus, if we aim to foster privacy engineering in practice, three – interdependent – main angles seem to be available: (1) increase user demand, (2) stricter and enforced obligations for  industry, or (3) provide easy to use, feasible and viable privacy-preserving technologies and methods. While addressing user demand is a topic we will not further consider here, the latter two approaches deserve more attention. Legislation already requires companies to apply privacy by design (e.g., DPbDD in Art. 25 GDPR). For multiple reasons, however, legislators usually refrain from stipulating specific technologies and methods. The actual technologies and methods to be used are to be derived from the state of the art, the risk caused by the processing, and other factors such as cost. A cutting-edge technology raising serious integration or operational cost will therefore not be considered obligatory to apply in most cases. The key to increased adoption of privacy engineering methods, tools, and technologies lies, thus, to a large extent on the supply side and, therefore, in the provision of easily usable, effective and economically viable artifacts. Only on the basis of widespread availability and adoption of these artifacts will recognized industry practices emerge to form the state of the art to be considered by controllers. 

The privacy engineering community’s best avenues to advancing the practical adoption of privacy engineering, thus, lies in consciously advancing this state of the art. This requires several steps: First, we – in academia and industry – have to provide concrete, sufficiently mature, and publicly available implementations to demonstrate feasibility and effectiveness and to introduce the respective mechanism to the practice. Second, we must ensure that the implementation can be integrated into realistic information systems with low effort and high protection efficiency (see developer-friendliness and low implementation efforts above), and third, we must demonstrate economic feasibility, i.e., that operational overheads are reasonable (typically through, e.g., performance experiments with realistic scenarios and payloads). 

Together, these three factors may then, depending on the specific cost-risk assessment for a particular use case, imply an implicit regulatory expectation to implement a privacy mechanism in practice. This “obligation through implementation” approach can be consciously applied for fostering the actual adoption of novel technical privacy mechanisms in real-world information systems engineering.

\section{What to take away}

Now that privacy engineering is gaining traction in industry, corporate Heads of Privacy Engineering, CISOs, and their teams need to be empowered with proper technical tools and methods. For this to happen, privacy engineering needs to better align with real-world information systems engineering. In this article we have argued that this requires several things. At a more technical layer, privacy engineering needs to broaden its view beyond mere anonymization, data minimization, and security, and needs to properly address “second-order non-functional properties” of privacy mechanisms, like re-usability or integration into established development practices. At a “policy” layer, it might be beneficial to abandon “all-or-nothing” approaches to privacy in some fields of academia in order to more easily bridge the gap between the academic world and industry. Regulators should continue striving towards risk-based approaches, while ensuring consistent, non-contradictory regulation. Companies, in turn, should seriously consider investing in their privacy engineering capabilities lest they find themselves lagging behind the state of the art by too far one day. 

\section*{Author Bios}

\begin{IEEEbiography}{Frank Pallas} is EXDIGIT professor of Privacy Engineering and Policy-Aligned Systems at the Paris Lodron University of Salzburg’s faculty for Digital and Analytical Sciences, Austria. At the time of writing, he was Senior Researcher at the Information Systems Engineering group of TU Berlin, Germany. He received is Ph.D. in computer science from TU Berlin, Germany, followed by senior research positions at the KIT’s Center for Applied Legal Studies and the FZI Forschungszentrum Informatik – both in Karlsruhe, Germany. Contact him at frank.pallas@plus.ac.at 
\end{IEEEbiography}

\begin{IEEEbiography}{Katharina Koerner} is a corporate development manager with Daiki, San Jose, CA 95129 USA. Her research interests include tech policy, privacy, security, and artificial intelligence regulation. Koerner received a Ph.D. in Euro- pean Law from Innsbruck Univer- sity, Austria. Contact her at kk@dai.ki.
\end{IEEEbiography}

\begin{IEEEbiography}{Isabel Barberá} is a Privacy Engineer, AI Advisor and the Co-founder of Rhite, a consultancy and research firm based in The Netherlands specialized in Responsible AI \& Privacy by Design. Isabel is also the author of PLOT4ai, an open-source tool for identifying and assessing AI risks using threat modeling methodology. Isabel is also one of the members of the ENISA ad hoc Working Group on Data Protection Engineering and a member of the EDPB Pool of Experts. She is a national expert contributing to the development of standards related to privacy engineering and AI at the ISO/CEN/CENELEC standardization bodies. Contact her at isabel@rhite.tech  
\end{IEEEbiography}

\begin{IEEEbiography}{Jaap-Henk Hoepman} is currently a guest professor at the PRISEC - Privacy And Security group of Karlstad University, Sweden. He is also an associate professor at the Digital Security group of the Radboud University, Nijmegen, the Netherlands, working for the iHub, the interdisciplinary research hub on Digitalization and Society. He is also an associate professor in IT Law section at the University of Groningen. He studies privacy by design and privacy friendly protocols, and is actively involved in the public debate concerning security and privacy in our society. In October 2021 his book Privacy Is Hard and Seven Other Myths. Achieving Privacy through Careful Design appeared at MIT Press. Contact him at jhh@cs.ru.nl 
\end{IEEEbiography}

\begin{IEEEbiography}{Meiko Jensen} is a Senior Lecturer for Cybersecurity and Privacy at Karlstad University, Sweden. His research interests are in the topics of privacy engineering, cybersecurity, cloud security and privacy, data protection by design, anonymity and pseudonymity, and protection goals of privacy. Meiko received his Ph.D. in the topic of cloud security from Ruhr University Bochum, Germany. Contact him at Meiko.Jensen@kau.se 
\end{IEEEbiography}

\begin{IEEEbiography}{Nandita Rao Narla} is the Head of Technical Privacy and Governance at DoorDash. Previously, she was a founding team member of a data profiling startup and held various leadership roles at EY, where she helped Fortune 500 companies build and mature privacy, cybersecurity, and data governance programs. She is a Senior Fellow at Future of Privacy Forum and serves on the Advisory Boards and technical standards committees for IAPP, Ethical Tech Project, X Reality Safety Initiative, Institute of Operational Privacy Design, and NIST. Contact her at nnarla@fpf.com 
\end{IEEEbiography}

\begin{IEEEbiography}{Nikita Samarin} is a privacy researcher and a doctoral candidate at the University of California, Berkeley. His research focuses on understanding the impact of existing software engineering practices on end-user privacy, and his record includes publications and papers on topics spanning mobile systems security, biometric authentication, and secure machine learning. Samarin received his BSc with Honors in Computer Science from the University of Edinburgh. Contact him at nsamarin@berkeley.edu.  
\end{IEEEbiography}

\begin{IEEEbiography}{Max-R. Ulbricht} is a technical officer at BlnBDI, the data protection authority of the Federal State of Berlin, Germany. Previously, he conducted research focusing on privacy-enhancing technologies at the Chair of Information Systems Engineering at the Technical University of Berlin (TU Berlin). Contact him at mru@meta-level.net 
\end{IEEEbiography}

\begin{IEEEbiography}{Isabel Wagner} is an Associate Professor in Cyber Security at the University of Basel, Switzerland. Her research interests are privacy and privacy-enhancing technologies, particularly metrics to quantify the effectiveness of privacy protection mechanisms, privacy protections for smart technologies, and measurement studies to create transparency for web and IoT systems. Her 2022 book, Auditing Corporate Surveillance Systems: Research Methods for Greater Transparency, is available from Cambridge University Press. Contact her at isabel.wagner@unibas.ch 
\end{IEEEbiography}

\begin{IEEEbiography}{Kim Wuyts} is a Cyber \& Privacy Manager at PwC Belgium with over 15 years of experience in privacy and security engineering. During her time as senior researcher at KU Leuven, Kim  led the development and extension of LINDDUN, a popular privacy threat modeling framework. Her mission is to raise privacy awareness and get organizations to embrace privacy and security best practices. Kim is a guest lecturer, and a public speaker at international privacy and security conferences such as RSA, OWASP Global AppSec, CPDP, and IAPP DPC. She is also a co-author of the Threat Modeling Manifesto, program co-chair of the International Workshop on Privacy Engineering (IWPE), and a member of ENISA’s working group on Data Protection Engineering. Contact her at kim.wuyts@pwc.com 
\end{IEEEbiography}

\begin{IEEEbiography}{Christian Zimmermann} is a senior expert for cybersecurity and privacy engineering at Robert Bosch GmbH. At Bosch Mobility, he is working on product cybersecurity strategy and enablement. Before he joined Bosch Mobility’s central Systems Engineering \& Technical Strategy department, he was a security and privacy researcher at Bosch Research. His research interests revolve around automotive security and privacy, transparency-enhancing technologies, privacy economics and privacy-preserving technologies. Christian received his Ph.D. from University of Freiburg and is a member of the ENISA Ad Hoc Working Group on Data Protection Engineering. Contact him at christian.zimmermann3@de.bosch.com 
\end{IEEEbiography}

\end{document}